\documentclass[acmsmall]{acmart}
\usepackage{colortbl}
\usepackage{xcolor}
\usepackage{subcaption}
\usepackage{caption}
\usepackage{soul,xcolor}
\usepackage{makecell}
\usepackage{wrapfig}
\usepackage{booktabs}

\definecolor{daricia}{cmyk}{0, 0.7808, 0.4429, 0.1412}
\AtBeginDocument{%
  \providecommand\BibTeX{{%
    \normalfont B\kern-0.5em{\scshape i\kern-0.25em b}\kern-0.8em\TeX}}}

\setcopyright{acmlicensed}
\acmJournal{PACMHCI}
\acmYear{2022} \acmVolume{6} \acmNumber{CSCW2} \acmArticle{355} \acmMonth{11} \acmPrice{15.00}\acmDOI{10.1145/3555775}

\usepackage{graphicx}
\usepackage{xcolor,colortbl}
\definecolor{high}{RGB}{205,205,205}
\definecolor{middle}{RGB}{225,225,225}
\definecolor{low}{RGB}{240,240,240}
\newcommand{\highcolor}{\cellcolor{high}} 
\newcommand{\midcolor}{\cellcolor{middle}} 
\newcommand{\lowcolor}{\cellcolor{low}} 

\newcommand\redsout{\bgroup\markoverwith{\textcolor{red}{\rule[0.5ex]{2pt}{0.4pt}}}\ULon}

\begin{document}

\title{Women's Perspectives on Harm and Justice after Online Harassment}

\author{Jane Im}
\affiliation{%
  \institution{University of Michigan}
  \country{USA}}
\author{Sarita Schoenebeck}
\affiliation{%
  \institution{University of Michigan}
  \country{USA}}
\author{Marilyn Iriarte}
\affiliation{%
  \institution{University of Maryland}
  \country{USA}}
\author{Gabriel Grill}
\affiliation{%
  \institution{University of Michigan}
  \country{USA}}
\author{Daricia Wilkinson}
\affiliation{%
  \institution{Clemson University}
  \country{USA}}
\author{Amna Batool}
\affiliation{%
  \institution{University of Michigan}
  \country{USA}}
\author{Rahaf Alharbi}
\affiliation{%
  \institution{University of Michigan}
  \country{USA}}
\author{Audrey Funwie}
\affiliation{%
  \institution{University of Michigan}
  \country{USA}}
\author{Tergel Gankhuu}
\affiliation{%
  \institution{University of Michigan}
  \country{USA}}
\author{Eric Gilbert}
\affiliation{%
  \institution{University of Michigan}
  \country{USA}}
\author{Mustafa Naseem}
\affiliation{%
  \institution{University of Michigan}
  \country{USA}}

\renewcommand{\shortauthors}{Jane Im et al.}

\begin{abstract}
Social media platforms aspire to create online experiences where users can participate safely and equitably. However, women around the world experience widespread online harassment, including insults, stalking, aggression, threats, and non-consensual sharing of sexual photos. This article describes women's perceptions of harm associated with online harassment and preferred platform responses to that harm. We conducted a survey in 14 geographic regions around the world (N = 3,993), focusing on regions whose perspectives have been insufficiently elevated in social media governance decisions (e.g. Mongolia, Cameroon). {Results show} that, on average, women perceive greater harm associated with online harassment than men, especially for non-consensual image sharing. Women also prefer most platform responses compared to men, especially removing content and banning users; however, women are less favorable towards payment as a response. Addressing global gender-based violence online requires understanding how women experience online harms and how they wish for it to be addressed. This is especially important given that the people who build and govern technology are not typically those who are most likely to experience online harms. 
\end{abstract}



\maketitle

\section{Introduction}
Online violence against women is a global problem. Women are subjected to abuse, harassment, stalking, threats, and misogyny on social media, and these experiences are increasingly documented in regions around the world \cite{heise1994gender}. A study by Amnesty International across 8 countries suggests that 23\% of women have experienced online harassment, and among those, 41\% felt that their physical safety was threatened \cite{amnesty2017}. A study by UNESCO similarly indicated that 73\% of women had experienced or observed some form of online violence \cite{unesco2015}. Human rights-based organizations including the United Nations, UNESCO, Human Rights Watch, and others have called for action to address and reduce online violence against women \cite{unesco-onlineviolence2021,humanrights-womenviolence-2019,aziz2015eliminating,un-takefive-2020}. 

Yet, social media companies have struggled to effectively govern online harassment. Governance on prominent platforms, such as WeChat, TikTok, and Instagram, relies on a complex and evolving set of policies to determine what content violates community guidelines and what does not \cite{gillespie2018custodians,jhaver2018online,ruan2016one}. These policies are enacted via a combination of algorithms and human content moderators who detect and process harmful content. However, these processes are imperfect and harmful content can remain on the platform while {appropriate} content is incorrectly removed \cite{gillespie2018custodians, roberts2019behind}. Further, governance processes have {insufficiently attended to} the disproportionate abuse experienced by marginalized and oppressed groups, such as women, people of color, sex workers, dissidents, transgender people, and disabled people \cite{marwick2018drinking,walker2020more,sambasivan2019they,vitak2017identifying, sultana2021unmochon,barwulor2021disadvantaged,mcdonald2021s}. 

One reason for these disparities is {that} platforms embrace principles of fairness that seek to minimize bias in both algorithmic and human moderation processes \cite{gillespie2018custodians}; however, treating content and users as equal entities falsely presumes that their underlying identities and experiences are equitable \cite{hoffmann2019fairness}. Instead, existing inequities like racism, misogyny, {and xenophobia} are perpetuated and magnified online, disproportionately harming those groups \cite{nakamura2008digitizing, kolko2013race, daniels2009cyber,jinsook2021resurgence}. While the systematic and structural inequalities experienced online long predate social media, the rise of social media platforms has enabled those inequalities to spread \cite{tynes2019race, packin2020disability, relia2019race, wu2018exploring}. It is not surprising then, that women continue to experience gender-based harassment including stalking, abuse, misogyny, unsolicited sexual photos (“dick pics”), and non-consensual sexual photo sharing (“revenge porn”) with little recourse or remedy \cite{citron2014criminalizing,vitis2017dick}.  

 This work builds on a growing interest among social media scholars to center the needs of harassment targets rather than the needs of the perpetrators or the platforms \cite{blackwell2017classification, schoenebeck2020drawing,bridy2018remediating,douek2020limits,santaclara,im2021yes}. Current models of governance typically rely on removing content that violates guidelines or temporarily or permanently banning perpetrators \cite{jhaver2019did,chandrasekharan2017you}. However, scholars have critiqued companies' use {of these} models because they overlook the experiences of those who are targeted by the harassment, forgoing an opportunity for them to receive just processes or outcomes {\cite{schoenebeck2020drawing}. Our work aims to understand how platform responses and governance models should be designed to center those who are {most} affected by online harms.}

Further, while gender-based violence is documented in nearly every region and culture around the world---and increasingly online as well---platform governance and policy-making has been largely conducted by relatively few Western perspectives, and these policies are then applied to regions and communities that are vastly different \cite{york2021silicon,barwulor2021disadvantaged}. This work builds on a growing chorus of scholars and activists across communities who are bringing voice and advocacy to women's experiences online outside of Western perspectives (e.g.  \cite{kumar2019engaging,sultana2018design,sambasivan2019toward,sambasivan2019they,york2021silicon}). 
Though focusing on women's experiences was not the primary goal at the outset of the project, our analyses indicated that differences between men and women's responses were strong predictors of harms and responses so we focused on those differences. Thus, this article is motivated by the following goals:
\begin{itemize}
    \item To understand perception of harm associated with different types of online harassment
   \item To understand preferences for platform responses associated with different types of online harassment
    \item To identify similarities or differences between women's and men's perceptions and preferences associated with online harassment 
\end{itemize}

We conducted an online survey with nearly 4,000 participants across 14 regions around the world, with a focus on non-Western regions that are underrepresented in social media {scholarship}. In the survey, participants were {presented four different online harassment scenarios} and seven possible platform responses. Our analysis shows that perceptions of harm differ by gender, with women perceiving greater harm than men in almost all scenarios. We also find differences in preferences for platform responses---across harassment scenarios, women tend to prefer responses more than men, especially banning accounts, revealing identity of perpetrators, and labeling content. We discuss the complexity of theorizing and applying justice models in the context of gender-based harassment online and conclude with implications for social media platform design and governance.

\section{Related Work}
\subsection{Online Harassment and Harms}
Online harassment refers to a wide range of behaviors that are hosted and enabled by technology platforms. These include hate speech, non-consensual sharing of sexual photos, stalking, doxxing, name calling and insults, impersonation, and public shaming \cite{blackwell2017classification,lenhart2016online}.
While online harassment was historically depicted as an outlier or fringe behavior, abusive behavior has been endemic in online communities since their inception. Young adults, women, queer people, people of color, disabled people, and other groups are disproportionately affected by online harassment, particularly when those identities intersect \cite{york2021silicon,online2015,twittertoxic}. In the U.S., scholars like Jessie Daniels, Lisa Nakamura, Danielle Citron, and others have traced the persistence of racism and misogyny in online spaces for decades, linking those behaviors to underlying offline social experiences \cite{kolko2013race, daniels2009cyber, nakamura2008digitizing}. 

The effects of harassment vary and can include anxiety, stress, fear, humiliation, self-blame, anger, and illness. Online harassment in particular can cause long-term damage due to the persistence and searchability of content. It also has a chilling effect on future disclosures; one study in the United States found that 27\% of Internet users were self-censoring their online posts due to fear of harassment \cite{lenhart2016online}. At an extreme, voices are silenced through threats of, or actual, violence and death. Qandeel Baloch, a social media icon in Pakistan, received abuse and harassment for her online persona that resulted in her brother murdering her as an ``honor killing'' \cite{QandeelBaloch}. In South Korea, where the rise of misogyny is recognized as a social issue \cite{jinsook2021resurgence}, celebrities Hara Goo and Sulli (Jin-ri Choi) died by suicide after experiencing large-scale cyberbullying and online harassment \cite{goo2019}.

Although harassment is instantiated online, targets of online harassment frequently report disruptions to their offline lives, including emotional and physical distress, changes to technology use or privacy behaviors, and increased safety and privacy concerns \cite{duggan2017online}. Some types of online harassment aim to disrupt a target’s offline life, such as swatting (i.e., reporting a crime to induce law enforcement agencies to investigate a target’s home) \cite{bernstein2016investigating}. Targets often choose to temporarily or permanently abstain from social media sites, despite the resulting isolation from information resources and support networks \cite{goldberg2019nobody}. Online harassment can also be disruptive to personal responsibilities, work obligations, and sleep due to the labor of reporting harassment to social media platforms or monitoring accounts for activity \cite{pittaro2007cyber,griffiths2002occupational}. 

{Though there are emerging efforts to categorize online harms broadly \cite{jiang2021understanding}, there are not yet standard frameworks for measuring harms associated with online harassment. Harms vary by domain (e.g., medical, environmental), by type of harm, and by severity of harm. The United Nations Declaration on the Elimination of Violence against Women identifies three overarching categories of harm: physical, sexual, and psychological \cite{unwomen}. These three categories are used widely across disciplines and industries and can intersect with other types of harm.} Physical harm involves direct bodily injury, or changes in environments that can relate to bodily safety, such as lack of access to housing. Sexual harm relates to sexual abuse, including rape. Psychological harm involves mental and emotional states, and is likely to co-occur with the other two categories; that is, it is unlikely a person could experience physical or sexual harm without also experiencing psychological harm. Other types of harm include financial harm, economic harm, and reproductive harm, though these are not the focus of this paper. Relational harm refers to interpersonal harm experienced by two people or a group of people, such as a family. {Family harm can vary from parental neglect to spousal abuse to isolation and abandonment. In the case of online harassment, it can manifest as shame from family members,} a phenomenon that has been documented in South Asian regions and may occur in other regions as well \cite{sultana2021unmochon,younas2020patriarchy,naseem2020designing,sultana2018design}.

\subsection{Platform Responses to Online Harassment}
Social media companies have come under increasing criticism for their failures to effectively govern online content \cite{gillespie2018custodians,roberts2019behind}. Companies maintain policies to determine what content can or cannot be on their platforms, and enforce those policies through a combination of algorithms and human workers. However, algorithms are crude hammers applied to complex social conflicts; they cannot understand the nuances of social conversations and contexts. As a result, harmful content persists and sometimes even thrives while innocuous content is incorrectly removed \cite{hosseini2017deceiving,gray2020intersectional}. Algorithmic regulation may also magnify harms against marginalized groups (e.g. by removing content that is combating racism while leaving up racist content). Although many of the inequities experienced online are not unique to the Internet, they can be magnified and exacerbated by social media features like ``likes'', follows, and algorithmic news feeds \cite{lewis2018alternative}. Content moderation is also done by human workers who are typically outsourced third-party contractors. These workers are paid low wages for rapid review of content that can be traumatic to look at \cite{roberts2019behind}. 

Social media companies have often adopted a ``neutral'' approach to governance that effectively absolves them of the responsibility to adjudicate harm \cite{gillespie2018custodians}. However, neutrality is not possible when platforms arbitrate millions of personal, nuanced, and contextualized posts daily. Additionally, content moderation decisions are not transparent to users, allowing sites to disguise the power they wield over the process \cite{roberts2019behind,gillespie2010politics}. While most people feel social media companies have a responsibility to remove offensive content from their platforms, few have confidence in companies to determine what offensive content should be removed \cite{laloggia2019us}. 

One proposal for moving forward is to expand beyond only content removal to consider other kinds of remedies \cite{schoenebeck2020drawing, goldman2021content, douek2020governing}. Sultana et al. explore the idea of a shame-based model of justice for women in the global south, noting that it is necessary in the absence of social and political support for women \cite{sultana2021unmochon}. In the United States, Schoenebeck et al. \cite{schoenebeck2020drawing} explore {whether punitive or restorative approaches} justice are desired among social media users. Hasinoff, Gibson, and Salehi have proposed that restorative justice approaches focused on repairing harms rather than punishment can improve content moderation \cite{hasinoff2020promise}. Though not explicitly oriented around justice frameworks, {Patel} et al. describe user-based (e.g. self-control, self-care) and platform-based (e.g., voting systems, reward systems) responses for detecting toxicity and promoting positivity on platforms \cite{patel2021user}.

Building on these collective directions, we explore what responses to online harassment are desirable for participants. We frame approaches like removing content or banning users as punitive models that echo criminal justice systems based on removal from society \cite{schoenebeck2020drawing}. We also focus on public shaming, such as publicly revealing a perpetrator's identity, given its prominence on social media today \cite{schoenebeck2020drawing,sultana2021unmochon,klonick2015re,ronson2016so}. Restorative justice models advocate for repairing harm so we ask survey participants about apologies, while concepts like reparation and racial justice {can relate to} compensation, so we ask about payment. Our intent is not to be comprehensive, which is unrealistic, but to imagine a range of possibilities for justice frameworks online. 

\subsection{Online Harassment and Gender}
{A}s Internet access has become available in many regions throughout the world, government, NGO, and industry statistics suggest that online abuse---often based on identity---is pervasive. In Mexico, women receive more online abuse than men including defamatory messages and contact through fake accounts \cite{mexico-digitalviolence}. In Mongolia, the majority of the population uses social media daily and although research is limited, hate speech and harassment appear to be widespread \cite{mongolia}. Online harassment is observed in Cameroon, too, despite lower social media adoption rates, and women have protested about online and offline sexual harassment they experience \cite{cameroon-stop,digital2019}. In South Korea, a massive online sex trafficking on Telegram called the ``Nth room case'' was discovered in 2020, which included minors' photos being sent to a quarter million users \cite{yun2020feminist,nthroom,joohee2021nth}. The Digital Rights Foundation in Pakistan launched a Cyber Harassment Helpline as part of its efforts to support online freedom of expression and the right to privacy for ``women, minorities and dissidents'' \cite{digitalrights}. 

{Despite this global prevalence}, Facebook itself has indicated that it focuses on priority areas and areas of high prevalence first and foremost \cite{facebook-loophole}; while it took about 7 days to address employee-reported inauthentic behavior on Facebook for content in the United States, it took 360 days to do the same for content in Mexico. In her book, \emph{Silicon Values}, Jillian York critiques social media platforms for their ``signals of mainstream, centralized American media, whose interests lie first and foremost in US affairs---and not just US affairs, but the things that matter most for the country's elites, who are still overwhelmingly white'' \cite{york2021silicon}. Indeed, most regulatory conversations involve an elite few leaders of Silicon Valley or west-coast-based companies who have outsized power over social media as well as its regulation \cite{roberts2019behind}. York explains how the United States public finally became concerned about the power of a concentrated few during the Trump presidency, a concern that has been well-known by activists and scholars globally for a decade or more \cite{york2021silicon}. 

This current work is inspired and motivated by efforts to move from a universal perspective to a more ``pluriversal'' perspective \cite{wong2020decolonizing}. As authors of a CSCW workshop on decolonizing argue, the pluriversal perspective seeks ``to foster `a world of many worlds' where contradicting ontologies and epistemologies can co-exist without needing to align with each other or claiming more validity over others” \cite{wong2020decolonizing}. Our work does not fit into decolonialist frameworks which place an emphasis in the geo-political as well as the body-political orientations when conducting research \cite{ali2016brief}. Nevertheless, our work aims to contribute to the goals of “de-centering” dominant assumptions about who governs social media and how they do so. 

The current work builds on movements that have been contesting the US-centered narrative \cite{sambasivan2019they, sambasivan2019toward, sultana2018design, sultana2021unmochon, kumar2020encountering,dye2016early}. Sambasivan et al. note that women in South Asian countries report experiences of cyberstalking, impersonation, and personal content leakages that cause emotional harm, reputation harm, romantic coercion, and domestic violence \cite{sambasivan2019they}. Women in South Asian regions tend to seek help from friends and family rather than formal channels, though family may not support targets of harassment \cite{sultana2021unmochon,sultana2018design}. {Jiang et al. \cite{jiang2021understanding} researched how participants from eight countries with the most Facebook users perceive various types of harm associated with online behavior more broadly} {(e.g., drug use; mutilation). They found that sexualization of minors was highest in harm across countries while spam was lowest, and patterns varied by country.} In the domain of content moderation, extensive non-Western work is driven by scholar-activists such as Dia Kayyali, Rasha Abdulla, Nanjira Sambuli, and many others who have raised the alarm about the colossal power of technology companies to decide what speech is allowed or not \cite{york2021silicon}. Of most concern, they note, are the risks that power brings for freedom of expression, as well as for documentation of human rights violations. 




\section{Methods}
{To explore online harassment harm and responses, we designed an online survey and recruited participants from 14 regions around the world.} {This section} describes survey design and translation, participant recruitment, participants demographics, and data cleaning {and analysis}.  

\subsection{Survey Design} \label{survey-design}
We designed the online survey iteratively as a team over a roughly four-month period {in 2020}. During the survey design phase, we discussed the survey design and brainstormed questions and topics in English. In some cases where multiple members of the research team spoke a {shared, non-English} language (e.g., Spanish, Urdu), they discussed the questions in their primary languages. We developed and revised the survey numerous times as members of the research team evaluated whether the questions would make sense in the culture of each region being studied. For the regions where the survey was translated into other languages, {the research team} also evaluated whether the questions still made sense in those languages. Eleven of the 14 regions were represented by members of the research team during the survey design phase; {research} collaborators from Austria, the Caribbean, and Saudi Arabia were added later. Where questions or wording did not make sense, we iteratively redesigned questions while checking that updated versions would continue to work in the other regions.

To aim for robust translation processes, we hired online translation services (with human translators) to translate the survey from English to the languages used in the surveys. A member of the research team who was bilingual in both the language used in the survey and in English also conducted an independent translation, and then members of the research team compared the independent translations with the paid translations to develop a single version. Each member of the research team then pilot tested the survey with a convenience sample of 2-4 people (typically family and friends) who spoke the relevant language. We used those pilot tests to refine and check translations. We administered the survey in the dominant local language of that region (see Table \ref{table:participant-demographics}). In India the survey was available in English and Hindi. In Cameroon the survey was available in English because our research team member was from Anglophone Cameroon (not Francophone Cameroon). 

{To develop the survey questions, w}e brainstormed varied types of online harassment and their contours---the severity of the harassment, who is affected, and the longevity of the harassment. We started with the six harassment types in Online Harassment Reports from Pew Research {(e.g. ``Has someone try to purposefully embarrass you'')} \cite{duggan2017online} as well as a broad literature review spanning work on content moderation (e.g., \cite{seering2018applications}), non-consensual image sharing (e.g. \cite{citron2014criminalizing, goldberg2019nobody}), non-Western gendered experiences (e.g. \cite{sambasivan2019they, sultana2018design}), and other areas. We selected a subset of harassment scenarios based on diversity of type of harassment, variance in severity of harm, and likelihood of being broadly relevant globally. We also focused on harassment scenarios that may have uncertain and inconsistent conceptualizations. For example, while non-consensual image sharing may seem obviously wrong to some people, others have claimed that if a person consensually took sexual photos, it implies them also consenting to those photos to be shared online \cite{citron2014criminalizing}. Our final survey used four harassment scenarios to minimize participant fatigue. 

In the first part of the survey, participants were presented with the the four harassment scenarios (see Table \ref{table:survey-definition}): Imagine a person has 1) taken sexual photos of you without your permission and shared them on social media (\textit{sexual photos}); 2) spread malicious rumors about you on social media (\textit{spread rumors}); 3) created fake accounts and sent you malicious comments through direct messages on social media (\textit{malicious messages}); and 4) insulted or disrespected you on social media (\textit{insulted or disrespected}). 

\begin{table}[t!]
\resizebox{0.9\textwidth}{!}
{
\begin{tabular}{l l}
\textbf{Harassment scenario} &  "Imagine a person has [...]"\\
\addlinespace[0.06cm]
sexual photos &
\makecell[l]{“taken sexual photos of you without your permission and \\ shared them on social media.”}\\
\addlinespace[0.07cm]
spread rumors & 
\makecell[l]{“spread malicious rumors about you on social media.”\\}\\
\addlinespace[0.07cm]
malicious messages & 
\makecell[l]{“created fake accounts and sent you malicious comments \\ through direct messages on social media.”}\\
\addlinespace[0.07cm]
insulted or disrespected &
“insulted or disrespected you on social media.”\\
\addlinespace[0.06cm]
\midrule
\addlinespace[0.06cm]
\textbf{Perceived harms} & \\
\addlinespace[0.06cm]
psychological harm &
“Would you be concerned for your psychological wellbeing?”\\
\addlinespace[0.06cm]
personal safety & 
“Would you be concerned for your personal safety?”\\
\addlinespace[0.1cm]
family reputation & 
“Would you be concerned for your family reputation?”\\
\addlinespace[0.06cm]
sexual harassment &
“Would you consider this sexual harassment against you?”\\\addlinespace[0.1cm]
\midrule
\addlinespace[0.06cm]
\textbf{Platform responses} & "The social media sites responds by [...]"\\
\addlinespace[0.06cm]
removing & 
“removing the content from the site.”\\
\addlinespace[0.06cm]
labeling &
“labeling the content as a violation of the site’s rules.”\\
\addlinespace[0.06cm]
banning &
“banning the person from the site.”\\
\addlinespace[0.06cm]
paying &
“paying you money.”\\
\addlinespace[0.06cm]
apology &
“requiring a public apology from the person.”\\
\addlinespace[0.06cm]
revealing &
“revealing the person’s real name and photograph publicly on the site.”\\
\addlinespace[0.06cm]
rating &
“by giving a negative rating to the person.”
\\ 
\end{tabular}
}
\vspace{9px}
\caption{Harassment scenarios, perceived harm, and platform responses.}
\vspace{-15px}
\label{table:survey-definition}
\end{table}

We adapted measures from prior literature to develop four measures associated with harm. For each scenario, participants were asked whether they would be concerned about the following: psychological harm, personal safety, family reputation, and whether they considered the scenario to be sexual harassment. In these adaptations we prioritized interpretability across languages, where everyday people would be likely to understand what we were asking. Thus, they deviate from English-language measures. Specifically, we used ``sexual harassment'' to capture the sexual nature of the harm, but chose not to use the term ``sexual harm'' because our pilot testing indicated participants were confused by translations of sexual harm across multiple languages. Sexual harassment is a broad term, like sexual harm, and can refer to both physical and psychological sexual behaviors. Similarly, we used the term ``personal safety'' as an alternative to physical harm, which also did not translate readily to other languages in our pilot testing.

{Perceived h}arm options were presented on a scale of ``Not at all concerned'' (coded as 1) to ``Extremely concerned'' (coded as 5) for psychological harm, personal safety, and family reputation and ``Definitely not'' (1) to ``Definitely'' (5) for sexual harassment. The sexual harassment item had a different set of options because {it did not pair well with the concerned anchors (“Would you be concerned for sexual harassment” and variations did not make sense)}. {We chose the Not at all concerned/Extremely concerned choice because it translated consistently across languages and because we could put it in the stem of the question to minimize the likelihood of acquiescence bias associated with Strongly Agree/Disagree measures \cite{saris2010comparing}}.

Then, participants were asked how desirable they found different types of responses to each scenario (see Table \ref{table:survey-definition}). {To develop the items for the platform responses, we similarly conducted a literature of content moderation practices and policies with a focus on proposed remedies \cite{schoenebeck2020drawing, hasinoff2020promise, sultana2021unmochon}. We chose banning, removing, and labeling as they are prominent examples of existing platform responses \cite{gillespie2018custodians}. We chose paying, apology, and revealing as they are proposed responses developed in prior work \cite{schoenebeck2020drawing}. We added rating because it has been discussed extensively in online communities literature \cite{chen2010social,kraut2012building} as a compelling approach to regulating online behavior.} {As with the harassment and harm measures, these are a sampling of possible responses but they are not comprehensive or representative.}

The final section contained social media use and demographic questions. The demographic questions were derived from Wave 6 of the World Values Survey (WVS) \cite{inglehart2014world} with some adaptations for use in an online survey (e.g., mobile usability). {We chose the WVS because it allowed us to ask single item questions with benchmarks in many countries and languages, which most validated scales do not provide.} The WVS is a long survey; we selected questions that captured demographic variables relevant to our project focus. This paper reports on questions related to gender (gender identity, marital status, and number of children) and gender equality (responses to: ``When jobs are scarce, men should have more rights to a job than women'' and ''When a mother works for pay, the children suffer.'').  

The question about participant gender identity expanded the WVS survey version (which includes "Male" and "Female" as options) to also include "Prefer to not disclose" and "Prefer to self-describe." This is aligned with recommendations to move beyond binary options \cite{scheuerman2020hci} though many scholars also recommend using "Man" and "Woman" instead of "Male" and "Female" which WVS and our survey did not do. Our survey did not ask about non-binary identity because {in some countries participants cannot safely identify as a gender outside of male or female}. The survey also did not ask about transgender identity or sexual identity, a limitation we return to in the discussion. Scheuerman et al., citing Meissner and Whyte, note that regions around the world have long histories during which gender was not binary which have been overwritten by racism and colonialism \cite{meissner2017theorizing}. 

\subsection{Participant Recruitment and Demographics}
We conducted an online survey with adult participants ages 18 {and over} from March 2020 through January 2021 across 22 countries: Austria, Cameroon, China, Colombia, India, South Korea, Malaysia, Mexico, Mongolia, Pakistan, Russia, Saudi Arabia, the United States, and nine countries within the Caribbean: St. Kitts and Nevis, Barbados, Dominica, St. Lucia, St. Vincent and the Grenadines, Jamaica, Grenada, Montserrat, and Antigua. This study was exempted from review by our institution’s Institutional Review Board. All participants completed a consent form. We used the survey company Cint to administer the survey in most of the regions in our sample. In the United States, we used the online recruitment platform Prolific. In two regions where Cint (and most global survey companies) have no presence (Caribbean countries, Mongolia), we recruited participants via word of mouth and snowball sampling. {The Caribbean and Mongolia samples are convenience samples and will be biased towards who was likely to see our research team members' invitations to participate. Cint and Prolific use a variety of guardrails to try to ensure diverse and robust survey participants; however, this is also a sample that will be biased towards Internet users and people who are likely to be on survey panels.}

Participants were compensated for their time through the survey company or directly, adjusted for exchange rates within the country and for time taken during a pilot test (e.g., Cameroon participants took longer than expected during the pilot survey so we increased their compensation). For the two regions where there was no panel presence, we compensated participants via mobile phone transfer in the local currency (Mongolian T{\"o}gr{\"o}g; East Caribbean Dollar). We aimed to pay a fair wage (e.g., \$15 USD/hour in the United States; 2000-3000 T{\"o}gr{\"o}g{/hour in Mongolia}). 

\begin{table*}[!t]

\centering
\resizebox{0.98\textwidth}{!}{
\begin{tabular}{ l c c c c c c c }

\textbf{Region} & \textbf{Language} & \vtop{\hbox{\strut \textbf{Number of}}\hbox{\strut \textbf{participants}}}  &  \textbf{Age} & \textbf{Men} & \textbf{Women} & \vtop{\hbox{\strut \textbf{Prefer not to}}\hbox{\strut \textbf{disclose}}} & \vtop{\hbox{\strut \textbf{Prefer to}}\hbox{\strut \textbf{self-describe}}} \\  \midrule
Austria & German & 251 & 37 & 50\%  & 48\% & 0\% & 2\%\\
Cameroon & English & 263 & 26 & 52\% & 45\% & 2\% & 1\% \\
Caribbean & English & 254 & 27 & 27\% & 69\% & 4\% & 0\%\\
China & Mandarin & 283\ & 36 & 50\% & 50\% & 0\% & 0\%\\
Colombia & Spanish (Colombian) & 296 & 34 & 47\% & 53\% & 0\% & 0\%\\
India & Hindi/English & 277 & 32 &  57\%  & 43\%  & 0\% & 0\%\\
South Korea & Korean & 252 & 41.5 & 47\% & 51\% & 2\% & 0\%\\
Malaysia & Malay & 298 & 34 & 47\% & 51\% & 1\% & 0\% \\
Mexico & Spanish (Mexican) & 306 & 33 & 51\% & 49\% & 0\% & 0\%\\
Mongolia & Mongolian & 367 & 21 & 32\% & 59\% & 8\% & 1\% \\
Pakistan & Urdu & 302 & 30 & 48\% & 50\% & 2\% & 0\% \\
Russia & Russian & 282 & 37 & 49\%  & 50\% & 0\% & 0\% \\
Saudi Arabia & Arabic & 258 & 33 & 55\% & 44\% & 2\% & 0\% \\
USA & English & 304 & 44 & 48\% & 51\% & 1\% & 1\%
\end{tabular}

}
\vspace{7px}
\caption{Regions studied, survey language, number of participants per region, average age, participant gender ratios. }
\vspace{-20px}
\label{table:participant-demographics}
\end{table*}

\subsubsection{Participant Demographics}  Women and men participated in similar rates across regions except for Caribbean countries (women: 69\%, men: 27\%, see Table \ref{table:participant-demographics}). The mean age was typically in the 30s, though Mongolia’s median was 21 while South Korea and United States’ median were each 41.5 and 44 (Table \ref{table:participant-demographics}). This pattern skews young but roughly reflects each country population, e.g., Mongolia’s median age is 28.2 years while South Korea and U.S medians are 43.7 and 38.3, respectively, according to the United Nations’ population estimates \cite{united20192019}. Participants’ self-reported income also diverged across regions, with participants in Austria reporting higher incomes and participants in Caribbean countries self-reporting lower incomes. More than half of the participants had education equivalent to a Bachelor degree for eight regions (Cameroon, China, Colombia, India, Malaysia, Russia, Saudi Arabia, United States); the other regions did not. Participants placed their political views as more ``left'' (1) than ``right'' (7) (mean of 3.22 with means ranging across countries from 2.8 in Austria to 3.9 in Korea; participants in Malaysia or Saudi Arabia did not receive this question because they may not be able to safely respond to it).

\subsection{Covid Impact Statement}
This study was conducted during the {beginning of the} global coronavirus pandemic. The pandemic inevitably impacted many or all of our participants’ lives given its global presence. We do not know how it may have affected their experiences or attitudes about online harassment. To benchmark our sample, we compared mean responses from our participants to mean responses from the World Values Survey. Because our sample and the WVS sample are different (e.g., recruited via online panels with questions optimized for mobile; recruited via door-to-door interviews with verbal question and answer choices) and the comparison is not the focus of the study, we qualitatively report differences. Our benchmarking was conducted with WVS Waves 6 and 7 (we selected Wave 7 when possible since it is more recent, but Wave 6 when questions or response choices were better aligned) for countries that are available in WVS data (China, Colombia, partial India, South Korea, Malaysia, Mexico, Pakistan, Russia, United States).  

Participants in our study reported better health than WVS participants. Participants in our study tended to have responses that were more aligned with gender equality for women {compared to WVS participants}---our participants more strongly disagreed that men should have more rights to a job than women, and disagreed that when a mother works for pay children suffer. This could be due to differences in people who are online versus those who are offline or due to response bias in our survey. It may also be because our sample trended about 5-10 years younger than the WVS samples (and than world population estimates). Our sample was much more likely to have spent money than to have saved it compared to WVS, which could relate to economic downturns in the first six months of Covid (when most data was collected), though it may also be that people who participate in online panels for compensation are looking for income and less likely to be in a position to save money. Participants' responses generally reflected expected trends from WVS data given known social, economic, and political differences (e.g., Pakistan and Saudi Arabia participants tended to have more conservative views about women working than other regions). 
 
\vspace{-7px}
\subsection{Data Cleaning and Analysis}

\textit{Data Cleaning.} We discarded low-quality responses based on duration of participation (using quantitative thresholds), quality of open-ended question responses (using subjective assessments of quality), and the number of unanswered multiple choice questions (i.e. more than 5 multiple choice questions skipped). Table \ref{table:participant-demographics} shows the final number of participants per region after data cleaning. We recruited participants from multiple Caribbean countries (Antigua and Barbuda, Barbados, Dominica, Grenada, Jamaica, Monserrat, St. Kitts and Nevis, St. Lucia, and St. Vincent). While each country of course has its own politics, culture, and economies, we felt that shared experiences across borders justified combining them for analysis. This was also a practical decision; Caribbean countries are small and we wanted to have a similar total sample size to other regions. One coauthor is from one of the Caribbean countries; the other countries are not represented in our research team.

\textit{Data Analysis.} We analyzed data using R software. We compared means between groups to report women's ratings of harm and justice-seeking responses. Then, we ran linear regressions to compare differences between women and men and to examine other demographic variables. {We used the Benjamini–Hochberg (BH) test to correct for multiple comparisons \cite{bretz2016multiple}.} For the means between groups, we ran Levene's tests to measure variance. For all cases, Levene's tests were significant (i.e., p-values were less than 0.05) indicating that homogeneity of variance assumption is violated. Thus, we used Welch one-way tests (with no assumption of equal variances) for nonparametric data and posthoc pairwise t-tests. The Welch one-way test can be appropriate when there is a sufficiently large sample size \cite{fagerland2012t}.

\vspace{-7px}
\subsection{Positionality Statement}
The project team included faculty, graduate students, and undergraduate students based at three universities in the United States. Despite the focus on centering non-Western views, the project is primarily centered at one institution in the United States and coauthors have affiliations with United States universities. Participation in the project included hourly paid research assistant positions, coauthorship, or both. Some project team members were active in the project through the survey design, data collection, and data analysis phases; others were active for only portions of the project. In our interest in decentering Western perspectives, we chose to collect data only in countries where a project team member could speak--as just one voice--to the culture and values of that country. While recognizing that one person does not reflect a country, for this project we decided to focus on regions that the research team collectively shared personal experiences with. As a result, the project team includes people from every country represented in the dataset. By ``from,'' we mean that they have a strong cultural affiliation with the country, including being born in and living in it throughout their childhood and/or early adulthood years. Many of the project team members were physically present in those countries during the project. 

\vspace{-7px}
\section{Results}
Results are presented in {two} sections: perceptions of harm associated with online harassment and preferred responses to online harassment. In each section, we present one visual representation of group means by gender. We use Welch tests to describe means among all participants, then use linear regressions to describe differences by gender and gender-related variables. 

\vspace{-2px}

\subsection{Perceptions of Harm of Online Harassment}

We conducted one-way Welch tests to compare means between the four harassment scenarios and the four harm measures. We compared group means between harassment scenarios and again between harm measures. Comparing between harassment scenarios tells us what types of harassment are perceived as more or less harmful (e.g. are rumors more harmful than insults?); comparing between harm types tell us what kinds of harassment might lead to that harm (e.g. is psychological harm or physical harm more prominent?).

Results were significant when comparing means across harassment scenarios for each harm measure: psychological harm, $F$(3,  8845.5) = 806.82, $p$ < 2.2e-16; personal safety, $F$(3, 8848.2) = 538.58, $p$ < 2.2e-16; family reputation, $F$(3,8824.1) = 709.52, $p$ < 2.2e-16; and sexual harassment, $F$(3, 8784.7) = 1321, $p$ < 2.2e-16. They were also significant when comparing means across harm measures for each harassment scenario: sexual photos, $F$(3,  8841.7) = 29.967, $p$ < 2.2e-16; spread rumors, $F$(3, 8848.3) = 110.6, $p$ < 2.2e-16; malicious messages, $F$(3, 8846.8) = 29.578, $p$ < 2.2e-16; insulted or disrespected, $F$(3, 8845.3) = 30.175, $p$ < 2.2e-16. 



\begin{wrapfigure}{r}{.55\textwidth}
\vspace{-8px}
\includegraphics[trim={0 0.25cm 0 0},clip, width=0.98\linewidth]{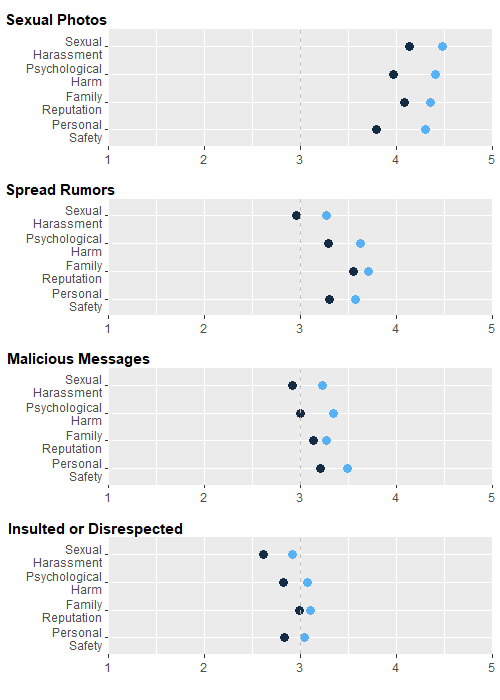}
\begin{center}
 \hspace{36px}\includegraphics[width=0.35\linewidth]{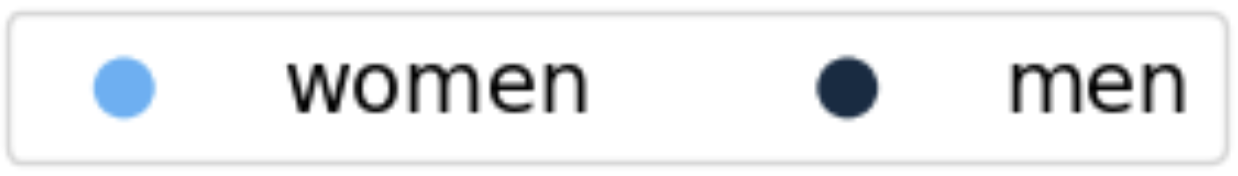}
\end{center}
  \vspace{-2px}
  \caption{Perceptions of harm by harassment scenario and harm type. Dark blue represents men's mean and light blue represents women's. 5 indicates the higher rating and 1 indicates lower.}
  \Description{Perceptions of harm by harassment scenario and harm type. Dark blue represents men and light blue represents women. 5 indicates the higher rating and 1 indicates lower.}
  \vspace{-10px}
  \label{figure:harmdiff}
\end{wrapfigure}

The \textit{sexual photos} scenario was rated as the most harmful scenario for all types of perceived harm, followed by \textit{spread rumors}, \textit{malicious messages}, and \textit{insulted or disrespected} scenarios (see patterns in Figure \ref{figure:harmdiff}). Harm measures varied, with sexual photos being rated highest as sexual harassment, spreading rumors being rated highest for family reputation, malicious messages being rated highest for personal safety, and insults or disrespect also rated highest for family reputation.

The post-hoc pairwise tests show that harassment scenarios were mostly different from each other in perceived harm (BH-adjusted $p$ < .05), except for sexual harassment in \textit{malicious messages} and \textit{spread rumors} scenarios {which were not significantly different from each other} (BH-adjusted $p$ = 0.11). {This is also seen in Figure \ref{figure:harmdiff}, which shows participants perceived a similar level of sexual harassment from the \emph{malicious messages} and \emph{spread rumors} scenarios.} 

Similarly, harm measures were mostly different from each other, with differences observed in five out of the six comparisons for each harassment scenario. Exceptions that were not significantly different from each other were personal safety and psychological harm for \textit{spread rumors}, family reputation and psychological harm for the \textit{sexual photos}, personal safety and psychological harm for \textit{insulted or disrespected}, and family reputation and psychological harm for \textit{malicious messages}. 


\begin{table}[b]
\begin{center}
\vspace{-10px}
\resizebox{.8\textwidth}{!}{%
\begin{tabular}{l l l l l}
\bf{Sexual Photos} & \bf{\makecell[l]{Psychological\\Harm}} & \bf{\makecell[l]{Personal \\Safety}} & \bf{\makecell[l]{Family \\Reputation}} & \bf{\makecell[l]{Sexual \\Harassment}} \\
\hline
Intercept             
& \highcolor $4.24 \; [ 4.10;  4.38]^{***}$ 
& \highcolor $3.80 \; [ 3.65; 3.94]^{***}$ 
& \highcolor $4.08 \; [ 3.94; 4.22]^{***}$ 
& \highcolor $4.81 \; [ 4.67;  4.94]^{***}$  \\
Woman                 
& \highcolor $0.41 \; [ 0.34;  0.48]^{***}$  
& \highcolor $0.51 \; [ 0.43; 0.58]^{***}$ 
& \highcolor $0.26 \; [ 0.18; 0.33]^{***}$ 
& \highcolor $0.29 \; [ 0.22;  0.36]^{***}$  \\
Gender undisclosed    
& \midcolor $0.42 \; [ 0.16;  0.69]^{**}$  
& \highcolor $0.70 \; [ 0.42; 0.98]^{***}$ 
& $0.14 \; [-0.12; 0.41]$     
& $-0.02 \; [-0.28;  0.24]$     \\
Gender self-described & $-0.35 \; [-0.93;  0.24]$     & $0.35 \; [-0.23; 0.93]$     & $-0.11 \; [-0.67; 0.46]$    & $-0.37 \; [-0.91;  0.18]$     \\
Women jobs            
& \midcolor $-0.13 \; [-0.22; -0.05]^{**}$ 
& $-0.09 \; [-0.18; 0.01]$    
& $-0.04 \; [-0.12; 0.05]$    
& \highcolor  $-0.37 \; [-0.45; -0.28]^{***}$ \\
Mother works          
& $-0.05 \; [-0.13;  0.04]$     
& \lowcolor $0.09 \; [ 0.01; 0.18]^{*}$ 
& $0.01 \; [-0.07; 0.10]$     
& \midcolor $-0.11 \; [-0.19; -0.03]^{**}$ \\
Is married            & $-0.00 \; [-0.09;  0.08]$     & $-0.08 \; [-0.16; 0.01]$    & $0.07 \; [-0.02; 0.15]$     & $-0.07 \; [-0.15;  0.01]$     \\
Has children          & $-0.01 \; [-0.04;  0.02]$     & $0.03 \; [-0.00; 0.06]$     & $0.02 \; [-0.01; 0.05]$     & $0.02 \; [-0.01;  0.05]$      \\
Adj. R$^2$            & $0.04$                        & $0.05$                      & $0.013$                      & $0.053$                        \\
\\
\end{tabular}
}

\resizebox{.8\textwidth}{!}{%
\begin{tabular}{l l l l l}
\bf{Spread Rumors} &  & & &  \\
\hline
Intercept             
& \highcolor $3.13 \; [ 2.97; 3.29]^{***}$ 
& \highcolor $3.03 \; [ 2.87; 3.19]^{***}$ 
& \highcolor $3.13 \; [ 2.97; 3.29]^{***}$ 
& \highcolor $2.76 \; [ 2.60; 2.92]^{***}$ \\
Woman                 
& \highcolor $0.35 \; [ 0.27; 0.43]^{***}$ 
& \highcolor $0.27 \; [ 0.19; 0.36]^{***}$ 
& \highcolor $0.18 \; [ 0.10; 0.27]^{***}$ 
& \highcolor $0.32 \; [ 0.24; 0.41]^{***}$ \\
Gender undisclosed    
& \lowcolor $0.39 \; [ 0.08; 0.70]^{*}$ 
& $0.29 \; [-0.02; 0.60]$    
& $0.12 \; [-0.19; 0.43]$     
& $0.26 \; [-0.04; 0.57]$     \\
Gender self-described & $0.46 \; [-0.19; 1.10]$     & $0.03 \; [-0.63; 0.68]$     & $0.02 \; [-0.63; 0.68]$     & $-0.18 \; [-0.84; 0.48]$    \\
Women jobs            & $0.07 \; [-0.03; 0.17]$     & $0.10 \; [-0.00; 0.20]$     
& \highcolor $0.18 \; [ 0.08; 0.28]^{***}$ & $-0.01 \; [-0.11; 0.09]$    \\
Mother works          & $0.01 \; [-0.09; 0.10]$     & $0.04 \; [-0.06; 0.13]$     & $0.05 \; [-0.05; 0.14]$     & $0.08 \; [-0.01; 0.18]$     \\
Is married            
& \lowcolor $0.12 \; [ 0.02; 0.21]^{*}$ & $0.04 \; [-0.06; 0.13]$     
& $0.08 \; [-0.01; 0.18]$     
& \highcolor $0.20 \; [ 0.11; 0.30]^{***}$ \\
Has children          & $0.00 \; [-0.03; 0.04]$     
& \highcolor $0.08 \; [ 0.05; 0.12]^{***}$ 
& \highcolor $0.07 \; [ 0.03; 0.10]^{***}$ & $0.02 \; [-0.02; 0.06]$     \\
Adj. R$^2$            & $.019$                      & $.018$                      & $.016$                      & $.022$                      \\
\\
\end{tabular}
}

\resizebox{.8\textwidth}{!}{%
\begin{tabular}{l l l l l}
\bf{Malicious Messages} &  & & &  \\
\hline
Intercept            
& \highcolor $2.80 \; [ 2.63; 2.97]^{***}$ 
& \highcolor $3.13 \; [ 2.96; 3.29]^{***}$ 
& \highcolor $2.60 \; [ 2.42; 2.78]^{***}$ 
& \highcolor $2.82 \; [ 2.66; 2.98]^{***}$ \\
Woman                 
& \highcolor $0.35 \; [ 0.26; 0.43]^{***}$ 
& \highcolor $0.26 \; [ 0.17; 0.35]^{***}$ 
& \highcolor $0.17 \; [ 0.08; 0.27]^{***}$ 
& \highcolor $0.30 \; [ 0.21; 0.38]^{***}$ \\
Gender undisclosed    & $0.31 \; [-0.01; 0.63]$     & $0.21 \; [-0.11; 0.53]$     & $0.27 \; [-0.06; 0.61]$     & $0.12 \; [-0.19; 0.43]$     \\
Gender self-described & $-0.54 \; [-1.22; 0.13]$    & $-0.47 \; [-1.14; 0.21]$    & $-0.62 \; [-1.32; 0.09]$    & $-0.33 \; [-0.99; 0.32]$    \\
Women jobs            & $0.06 \; [-0.05; 0.16]$     & $-0.03 \; [-0.13; 0.08]$    
& \midcolor $0.16 \; [ 0.05; 0.27]^{**}$ & $-0.09 \; [-0.20; 0.01]$    \\
Mother works          & $0.03 \; [-0.07; 0.13]$     & $0.03 \; [-0.07; 0.13]$     & $0.10 \; [-0.00; 0.21]$     & $0.09 \; [-0.00; 0.19]$     \\
Is married            & $0.03 \; [-0.07; 0.13]$     & $-0.03 \; [-0.13; 0.07]$    & $0.07 \; [-0.03; 0.18]$     & $0.10 
\; \lowcolor [ 0.01; 0.20]^{*}$ \\
Has children          
& \highcolor $0.08 \; [ 0.04; 0.11]^{***}$ 
& \highcolor $0.10 \; [ 0.06; 0.14]^{***}$ 
& \highcolor $0.13 \; [ 0.09; 0.17]^{***}$ 
& \highcolor $0.07 \; [ 0.03; 0.10]^{***}$ \\
Adj. R$^2$            & $0.021$                      & $0.017$                      & $0.026$                      & $0.021$                      \\
\\
\end{tabular}
}

\resizebox{.8\textwidth}{!}{%
\begin{tabular}{l l l l l}
\bf{Insult or Disrespect} & & & &  \\
\hline
Intercept             
& \highcolor $2.23 \; [ 2.06; 2.40]^{***}$ 
& \highcolor $2.23 \; [ 2.05; 2.40]^{***}$ 
& \highcolor $2.13 \; [ 1.95; 2.30]^{***}$ 
& \highcolor $2.27 \; [ 2.11; 2.43]^{***}$ \\
Woman                
& \highcolor $0.31 \; [ 0.22; 0.40]^{***}$ 
& \highcolor $0.26 \; [ 0.17; 0.35]^{***}$ 
& \highcolor $0.18 \; [ 0.09; 0.28]^{***}$ 
& \highcolor $0.34 \; [ 0.25; 0.42]^{***}$ \\
Gender undisclosed    & $0.23 \; [-0.09; 0.55]$     
&\lowcolor $0.43 \; [ 0.10; 0.76]^{*}$ 
&\lowcolor $0.37 \; [ 0.03; 0.71]^{*}$ & $0.14 \; [-0.17; 0.45]$     \\
Gender self-described & $-0.53 \; [-1.20; 0.14]$    & $-0.09 \; [-0.79; 0.60]$    & $-0.16 \; [-0.88; 0.55]$    & $0.34 \; [-0.31; 0.99]$     \\
Women jobs            
& \highcolor $0.29 \; [ 0.19; 0.40]^{***}$ 
& \highcolor $0.25 \; [ 0.14; 0.36]^{***}$ 
& \highcolor $0.36 \; [ 0.25; 0.47]^{***}$ 
& $0.05 \; [-0.05; 0.15]$     \\
Mother works          
& $0.07 \; [-0.03; 0.17]$     
& $0.09 \; [-0.01; 0.20]$     
& \midcolor $0.17 \; [ 0.06; 0.27]^{**}$ 
& \lowcolor $0.11 \;  [ 0.01; 0.20]^{*}$ \\
Is married            
& \highcolor $0.29 \; [ 0.19; 0.39]^{***}$ 
& \highcolor $0.20 \; [ 0.10; 0.31]^{***}$ 
& \highcolor $0.20 \; [ 0.09; 0.30]^{***}$ 
& \highcolor $0.24 \; [ 0.15; 0.34]^{***}$ \\
Has children          
& $-0.02 \; [-0.06; 0.02]$   
& \midcolor $0.05 \; [ 0.02; 0.09]^{**}$ 
& \midcolor $0.06 \; [ 0.02; 0.10]^{**}$ 
& $0.02 \; [-0.02; 0.05]$     \\
Adj. R$^2$            & $0.033$                      & $0.03$                      & $0.036$                      & $0.027$                      \\
\\
\end{tabular}
}
\vspace{5px}
\caption{Linear regression models for perceptions of harm showing coefficients and confidence intervals.}
\label{table:harm_regressionoutput}
    \vspace{-12px}
\end{center}
\end{table}

\subsubsection{Comparing women's perceptions of harm to men's} We fitted a series of linear regression models modeling harm as the dependent variable and demographic variables included in the survey as the independent variables (see Table \ref{table:harm_regressionoutput}, see visual representation in Figure \ref{figure:harmdiff}). We ran 16 models for the four harassment scenario and four harm type pairings. Variance inflation factors (VIF) were all less than 2 indicating multicollinearity was not an issue. 

Women perceived significantly higher harm than men. This pattern was observed in all 16 harassment scenario-harm pairings with gender accounting for relatively large variability in the models. Results for undisclosed or self-described gender were not significant for most scenario and harm pairings, however, these categories were underrepresented in our data (Table \ref{table:participant-demographics}). For the \textit{insulted or disrespected} scenario, being married was a predictor of increases in perception of harm for all four measures but this was not observed for other harassment types. For the \textit{malicious messages} scenario, having children was a predictor of increases in perception of harm for all four measures. This may be because sending comments through direct messages on social media is a fear parents often have for their children.


\subsubsection{Attitudes about gender and perceptions of harm.}
For three of the harassment scenarios (\textit{spread rumors}, \textit{insulted or disrespected}, \textit{malicious messages}), participants who tended to agree with the statement ``When jobs are scarce, men should have more rights to a job than women'' perceived greater harm to family reputation. This suggests that people who are less likely to support gender equality are also more concerned about family reputation after online harassment.
In contrast, for the \textit{sexual photos} scenario, people who tended to disagree with the statement rated those scenarios as higher in terms of psychological harm and sexual harassment.

\begin{wrapfigure}{r}{.55\textwidth}
    \vspace{-5px}
    \includegraphics[trim={0 0.25cm 0 0.25cm},clip,width=0.98\linewidth]{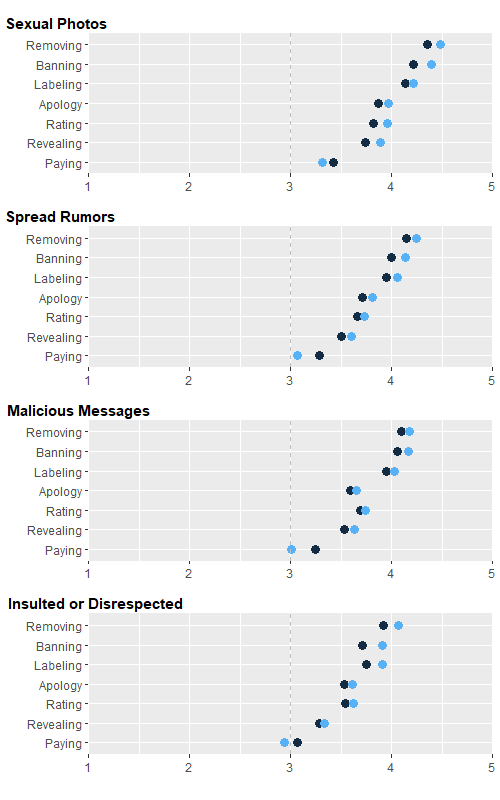}
    \begin{center}
    \hspace{29px}\includegraphics[width=0.35\linewidth]{legend.png}
    \end{center}
     \vspace{-2px}
    \caption{Preferences for responses by harassment scenario and response type. Dark blue represents men's mean and light blue represents women's. 5 indicates the higher rating and 1 indicates lower.}
    \Description{Preferences for responses by harassment scenario and response type. Dark blue represents men and light blue represents women. 5 indicates the higher rating and 1 indicates lower. }    
    \label{fig:remedy-diff-panel}
    \vspace{-7px}
\end{wrapfigure}

\subsection{Preferred Responses to Online Harassment}

In the prior section, we measured perceptions of harm for each of the four harassment scenarios. Here we want to understand which responses are preferred for each of the harassment scenarios. This allows us to answer the question, if someone experiences that type of online harassment, what response might they prefer? 
In the last section, we compared group means between harassment scenarios (one comparison for each of the four harm measures) and also between the harm measures (one comparison for each of the four harassment scenarios) because it was useful to be able to interpret both. Here we only compare group means between response types (for the four harassment scenarios) because it is primarily useful to know what responses are preferred for a given type of harassment.


One-way Welch tests for preferred responses were significant for all four harassment scenarios: \textit{sexual photos} scenario, $F$(6, 12381) = 304.87, $p$ < 2.2e-16; \textit{spread rumors} scenario, $F$(6, 12395) = 291.24, $p$ < 2.2e-16; \textit{malicious messages} scenario, $F$(6, 12399) = 307.57, $p$ < 2.2e-16; and \textit{insulted or disrespected} scenario, $F$(6, 12402) = 262.18, $p$ < 2.2e-16. 


As evident in Figure \ref{fig:remedy-diff-panel}, removing, banning, and labeling were the three most preferred responses. Payment was the least preferred in all four harassment scenarios. Posthoc pairwise t-tests showed that most ratings of responses differed significantly for most, though not all, pairings. The rating and apology responses tended not to be significantly different from each other. 


\subsubsection{Comparing women's preferred responses to men's}
We fitted a series of linear regression models for each of the four harassment scenarios and each of the seven response options (28 total, see Table \ref{table:response_regressionoutput}). Platform response options were modeled as the dependent variable and demographic data related to gender as the independent variables. Multicollinearity was not an issue as VIF were all less than 2. 
 

The difference in preferences between women and men was observed most frequently for banning: women prefer banning more than men in all four harassment scenarios. Women also preferred the apology response compared to men for all four harassment scenarios. This was followed by revealing which was preferred by women more than men in three of the four scenarios. Labeling and rating were preferred by women in two scenarios and removing was preferred in one scenario. However, one outlier is payment which was inverted; women preferred payment \textit{less} than men did in three of the harassment scenarios. 

\begin{table}[b!]
\begin{center}

\resizebox{\textwidth}{!}{
\begin{tabular}{l l l l l l l l}
\bf{\makecell[l]{Sexual \\Photos}}  & \bf{\makecell[l]{Removing}} & \bf{\makecell[l]{Labeling}} & \bf{\makecell[l]{Banning}} & \bf{\makecell[l]{Payment}} & \bf{\makecell[l]{Apology}} & \bf{\makecell[l]{Revealing}} & \bf{\makecell[l]{Rating}} \\
\hline
Intercept      
& \highcolor $5.05 \; [4.92; 5.17]^{***}$ 
& \highcolor $4.62 \; [ 4.47; 4.77]^{***}$ 
& \highcolor $4.86 \; [4.72; 5.00]^{***}$ 
& \highcolor $3.57 \; [ 3.38; 3.76]^{***}$ 
& \highcolor $3.91 \; [3.74; 4.08]^{***}$ 
& \highcolor $3.78 \; [ 3.61; 3.96]^{***}$ 
& \highcolor $4.02 \; [3.85; 4.19]^{***}$ \\
Woman          & $.06 \; [-.01;  .12]$      & $.04 \; [ -.04;  .11]$      & \highcolor $.13 \; [ .06;  .20]^{***}$  
& \lowcolor $-.11 \; [ -.21; -.01]^{*}$ 
& \lowcolor $.11 \; [ .02;  .20]^{*}$  
& \highcolor $.16 \; [  .07;  .26]^{***}$  
& \lowcolor $.11 \; [ .02;  .20]^{*}$  \\
Undisclosed    & $.08 \; [-.16;  .33]$      & $-.14 \; [ -.42;  .15]$     & $-.03 \; [-.29;  .24]$     
& \midcolor $-.52 \; [ -.89; -.15]^{**}$ 
& $.07 \; [-.26;  .40]$      
& $.03 \; [ -.30;  .37]$      
& $.28 \; [-.05;  .62]$      \\
Self-described 
& $-.26 \; [-.77;  .25]$     & $-.48 \; [-1.08;  .13]$     & $-.36 \; [-.91;  .20]$     & $-.41 \; [-1.19;  .36]$     & $-.17 \; [-.86;  .51]$     & $-.41 \; [-1.11;  .28]$     & $.15 \; [-.54;  .85]$      \\
Women jobs     
& \highcolor $-.39 \; [-.47; -.31]^{***}$ 
& \highcolor $-.27 \; [ -.37; -.18]^{***}$ 
& \highcolor $-.39 \; [-.48; -.31]^{***}$ 
& \midcolor $-.16 \; [ -.28; -.04]^{**}$ & $-.02 \; [-.12;  .09]$     & $-.08 \; [ -.19;  .03]$     
& \highcolor $-.19 \; [-.30; -.09]^{***}$ \\
Mother works   
& \midcolor $-.12 \; [-.20; -.05]^{**}$ 
& \lowcolor $-.11 \; [ -.20; -.03]^{*}$ 
& \midcolor $-.12 \; [-.20; -.03]^{**}$ & $-.07 \; [ -.18;  .04]$     
& $-.09 \; [-.19;  .01]$     
& $-.04 \; [ -.14;  .07]$    
& $-.04 \; [-.14;  .06]$     \\
Is married     
& \lowcolor $-.08 \; [-.16; -.01]^{*}$ & $-.04 \; [ -.13;  .05]$     & $.03 \; [-.05;  .11]$      & $.11 \; [ -.01;  .22]$      
&\lowcolor $.11 \; [ .01;  .21]^{*}$  
& \midcolor $.16 \; [  .06;  .27]^{**}$  & $-.02 \; [-.12;  .08]$     \\
Has children   
& \lowcolor $.03 \; [ .01;  .06]^{*}$  
& \midcolor  $.04 \; [  .01;  .08]^{**}$  
& $.02 \; [-.01;  .05]$     
& \highcolor $.08 \; [  .04;  .12]^{***}$  & $.02 \; [-.02;  .05]$      & $.03 \; [ -.01;  .07]$      
& \highcolor $.09 \; [ .05;  .13]^{***}$  \\
Adj. R$^2$     & $.04$                      & $.017$                       & $.037$                      & $.011$                       & $.003$                      & $.008$                       & $.012$                      \\
\\
\end{tabular}
}

\resizebox{\textwidth}{!}{
\begin{tabular}{l l l l l l l l}

\bf{\makecell[l]{Spread\\Rumors}} &  &  &  &  &  &  &  \\
\hline
Intercept      
& \highcolor $4.62 \; [ 4.48; 4.75]^{***}$ 
& \highcolor $4.32 \; [ 4.17; 4.47]^{***}$ 
& \highcolor $4.34 \; [ 4.19; 4.48]^{***}$ 
& \highcolor $3.11 \; [ 2.93; 3.29]^{***}$ 
& \highcolor $3.51 \; [ 3.35; 3.68]^{***}$ 
& \highcolor $3.20 \; [ 3.03; 3.37]^{***}$ 
& \highcolor $3.69 \; [3.53; 3.86]^{***}$ \\
Woman          
& $.07 \; [ -.01;  .14]$      
& \lowcolor $.08 \; [  .01;  .16]^{*}$  
& \midcolor $.12 \; [  .04;  .19]^{**}$  
& \highcolor $-.19 \; [ -.28; -.09]^{***}$ 
& \midcolor $.12 \; [  .03;  .20]^{**}$  
& \lowcolor $.12 \; [  .03;  .21]^{*}$  & $.06 \; [-.03;  .14]$      \\
Undisclosed    & $-.04 \; [ -.30;  .23]$     & $.02 \; [ -.26;  .30]$      & $.06 \; [ -.22;  .34]$      & $-.20 \; [ -.55;  .15]$     
& \lowcolor $.40 \; [  .08;  .73]^{*}$  & $.18 \; [ -.15;  .52]$      & $.27 \; [-.04;  .59]$      \\
Self-described & $-.50 \; [-1.05;  .06]$     & $-.42 \; [-1.01;  .16]$   & \lowcolor $-.61 \; [-1.19; -.03]^{*}$
& \lowcolor $-.74 \; [-1.47; -.01]^{*}$ & $-.62 \; [-1.29;  .05]$     & $-.63 \; [-1.33;  .06]$     & $-.14 \; [-.80;  .53]$     \\
Women jobs     
& \highcolor $-.28 \; [ -.37; -.20]^{***}$ 
& \highcolor $-.23 \; [ -.32; -.14]^{***}$ 
& \highcolor $-.30 \; [ -.39; -.20]^{***}$ 
& \lowcolor $-.11 \; [ -.23; -.00]^{*}$ 
& $.07 \; [ -.03;  .18]$      
& $.02 \; [ -.09;  .12]$      
& \lowcolor $-.12 \; [-.22; -.02]^{*}$ \\
Mother works   
& \lowcolor $-.11 \; [ -.19; -.02]^{*}$ 
& \midcolor $-.14 \; [ -.23; -.06]^{**}$ 
& $-.04 \; [ -.12;  .05]$     
&  \lowcolor $.11 \; [  .00;  .21]^{*}$  
& $-.04 \; [ -.14;  .06]$     
& $.06 \; [ -.04;  .16]$      
& $-.04 \; [-.14;  .05]$     \\
Is married     
& $-.03 \; [ -.11;  .05]$     
& $.07 \; [ -.02;  .16]$      
& $.05 \; [ -.03;  .14]$      
& $.08 \; [ -.03;  .19]$    
&\midcolor $.15 \; [  .05;  .25]^{**}$  
&\midcolor $.14 \; [  .04;  .24]^{**}$  & $.07 \; [-.03;  .17]$      \\
Has children   
&\highcolor  $.05 \; [  .02;  .08]^{***}$  
&\highcolor  $.07 \; [  .04;  .11]^{***}$  
&\highcolor  $.06 \; [  .03;  .10]^{***}$  
&\highcolor  $.10 \; [  .06;  .14]^{***}$  
&\lowcolor $.04 \; [  .01;  .08]^{*}$  
&\highcolor $.11 \; [  .07;  .15]^{***}$  
&\highcolor $.11 \; [ .07;  .15]^{***}$  \\
Adj. R$^2$     & $.022$                       & $.022$                       & $.022$                       & $.015$                       & $.01$                       & $.02$                       & $.015$                       \\
\\
\end{tabular}
}

\resizebox{\textwidth}{!}{
\begin{tabular}{l l l l l l l l}
\bf{\makecell[l]{Malicious \\Messages}}  &  &  &  &  &  &  &  \\
\hline
Intercept     
&\highcolor $4.56 \; [ 4.42; 4.70]^{***}$ 
&\highcolor $4.34 \; [4.20; 4.49]^{***}$ 
&\highcolor $4.49 \; [ 4.34; 4.63]^{***}$ 
&\highcolor $3.07 \; [2.89; 3.26]^{***}$ 
&\highcolor $3.33 \; [ 3.16; 3.50]^{***}$ 
&\highcolor $3.37 \; [ 3.20; 3.54]^{***}$ 
&\highcolor $3.81 \; [3.64; 3.97]^{***}$ \\
Woman          
& $.02 \; [ -.05;  .09]$     
& $.06 \; [-.02;  .14]$      
&\lowcolor $.09 \; [  .01;  .16]^{*}$  
&\highcolor $-.21 \; [-.30; -.11]^{***}$ 
&\lowcolor $.09 \; [  .00;  .18]^{*}$  
&\lowcolor $.11 \; [  .02;  .20]^{*}$  
& $.03 \; [-.06;  .11]$      \\
Undisclosed    
& $-.13 \; [ -.40;  .14]$     
& $-.10 \; [-.38;  .18]$     
& $-.04 \; [ -.32;  .23]$     
& \lowcolor $-.45 \; [-.80; -.09]^{*}$ 
& $.14 \; [ -.19;  .47]$      
& $-.27 \; [ -.59;  .05]$     
& $-.03 \; [-.34;  .29]$     \\
Self-described & $-.46 \; [-1.03;  .11]$     & $-.13 \; [-.72;  .46]$     & $-.44 \; [-1.01;  .13]$     & $-.22 \; [-.95;  .52]$     
&\lowcolor $-.74 \; [-1.42; -.05]^{*}$ 
&\midcolor $-.96 \; [-1.64; -.28]^{**}$ & $-.16 \; [-.82;  .51]$     \\
Women jobs     
&\highcolor $-.29 \; [ -.38; -.20]^{***}$ 
&\highcolor $-.22 \; [-.31; -.13]^{***}$ 
&\highcolor $-.27 \; [ -.36; -.18]^{***}$ 
& $-.10 \; [-.22;  .01]$     
& $.05 \; [ -.06;  .16]$      
& $-.01 \; [ -.11;  .10]$     
&\highcolor $-.19 \; [-.30; -.09]^{***}$ \\
Mother works   
&\lowcolor $-.09 \; [ -.18; -.01]^{*}$ 
&\highcolor $-.16 \; [-.25; -.07]^{***}$ 
&\lowcolor $-.09 \; [ -.17; -.01]^{*}$ 
& $.11 \; [-.00;  .21]$      
& $.02 \; [ -.08;  .12]$      
& $.01 \; [ -.09;  .11]$      
& $-.04 \; [-.13;  .06]$     \\
Is married     & $.03 \; [ -.05;  .11]$     
&\lowcolor $.09 \; [ .00;  .18]^{*}$  
& $.07 \; [ -.01;  .16]$      
&\lowcolor $.12 \; [ .01;  .23]^{*}$  
&\highcolor $.23 \; [  .13;  .33]^{***}$  
&\highcolor $.20 \; [  .10;  .30]^{***}$  
&\midcolor $.14 \; [ .04;  .23]^{**}$  \\
Has children   
&\midcolor $.05 \; [  .02;  .08]^{**}$  
&\midcolor $.05 \; [ .02;  .08]^{**}$ 
& $.02 \; [ -.01;  .05]$    
& \highcolor$.07 \; [ .03;  .12]^{***}$  
& $.03 \; [ -.01;  .06]$      
& \midcolor$.06 \; [  .02;  .10]^{**}$  
& \highcolor$.09 \; [ .06;  .13]^{***}$  \\
Adj. R$^2$     & $.019$                       & $.018$                      & $.018$                       & $.015$                      & $.011$                       & $.016$                       & $.017$                      \\
\\
\end{tabular}
}

\resizebox{\textwidth}{!}{
\begin{tabular}{l l l l l l l l}
\bf{\makecell[l]{Insulted or\\Disrespected}}  &  &  &  &  &  &  &  \\
\hline
Intercept      
&\highcolor $4.19 \; [ 4.04; 4.34]^{***}$ 
&\highcolor $3.94 \; [3.78; 4.09]^{***}$ 
&\highcolor $3.69 \; [ 3.54; 3.85]^{***}$ 
&\highcolor $2.73 \; [ 2.55; 2.92]^{***}$ 
&\highcolor $3.28 \; [3.11; 3.44]^{***}$ 
&\highcolor $2.84 \; [ 2.66; 3.01]^{***}$ 
&\highcolor $3.52 \; [3.35; 3.68]^{***}$ \\
Woman         
&\midcolor $.12 \; [  .04;  .20]^{**}$ 
&\highcolor $.15 \; [ .07;  .23]^{***}$  
&\highcolor $.21 \; [  .13;  .29]^{***}$ 
& $-.08 \; [ -.17;  .02]$     
&\midcolor $.12 \; [ .03;  .21]^{**}$  & $.09 \; [ -.00;  .18]$      
&\lowcolor $.10 \; [ .01;  .18]^{*}$  \\
Undisclosed    & $-.16 \; [ -.45;  .12]$     & $-.01 \; [-.30;  .29]$     & $.23 \; [ -.07;  .53]$      & $-.08 \; [ -.42;  .27]$     & $.26 \; [-.06;  .59]$      & $-.03 \; [ -.38;  .31]$     & $-.05 \; [-.37;  .26]$     \\
Self-described & $-.55 \; [-1.15;  .06]$     & $.00 \; [-.62;  .62]$      & $-.39 \; [-1.02;  .24]$     & $-.40 \; [-1.13;  .32]$     & $-.13 \; [-.81;  .55]$     & $-.35 \; [-1.06;  .37]$     & $-.12 \; [-.78;  .55]$     \\
Women jobs    
&\highcolor $-.20 \; [ -.29; -.10]^{***}$ 
&\highcolor $-.17 \; [-.26; -.07]^{***}$ 
&\lowcolor $-.10 \; [ -.20; -.00]^{*}$ & $.01 \; [ -.11;  .12]$   
&\lowcolor $.11 \; [ .00;  .21]^{*}$  
& $.06 \; [ -.06;  .17]$      
&\lowcolor $-.12 \; [-.22; -.01]^{*}$ \\
Mother works   & $-.07 \; [ -.16;  .02]$     & $-.08 \; [-.17;  .01]$     & $.01 \; [ -.08;  .10]$      & $.10 \; [ -.01;  .21]$      & $-.03 \; [-.13;  .06]$     
&\lowcolor $.11 \; [  .01;  .22]^{*}$  & $-.02 \; [-.12;  .08]$     \\
Is married     & $-.00 \; [ -.09;  .08]$     & $.08 \; [-.01;  .17]$     & \highcolor $.16 \; [  .06;  .25]^{***}$  
& \midcolor $.15 \; [  .04;  .26]^{**}$  
& \highcolor $.25 \; [ .15;  .35]^{***}$  
& \highcolor $.27 \; [  .16;  .37]^{***}$  
& \lowcolor $.12 \; [ .02;  .22]^{*}$  \\
Has children   
& \highcolor $.08 \; [  .04;  .11]^{***}$  
& \highcolor $.07 \; [ .03;  .10]^{***}$  
& \lowcolor $.04 \; [  .00;  .07]^{*}$  
& \highcolor $.07 \; [  .03;  .11]^{***}$  
& $.01 \; [-.03;  .05]$      
& \highcolor $.07 \; [  .03;  .11]^{***}$  
& \highcolor $.09 \; [ .06;  .13]^{***}$  \\
Adj. R$^2$     & $.016$                       & $.015$                      & $.014$                       & $.012$                       & $.011$                      & $.022$                       & $.015$                      \\
\\
\end{tabular}
}
\vspace{5px}

\caption{Linear regression models for preferences for responses showing coefficients and confidence intervals.}
\label{table:response_regressionoutput}

\end{center}
\end{table}

Participants who tended to agree with the statement ``When jobs are scarce, men should have more rights to a job than women'' were less likely to prefer responses for 18 out of the 28 pairings. In other words, people who tended not to support gender equality were also less supportive of many responses to online harassment. 
Interestingly, participants who had children preferred 22 of the 28 responses. This likely means that participants who are parents were thinking about their children when responding to the questions even if the questions were directed to the parent.

\subsection{Limitations when interpreting results}
This work used surveys to capture people's perceptions on online harassment and various approaches to harm online. We chose multiple-choice questions as our goal was to compare responses from a large sample of participants and regions. However, future research could use qualitative methods to better understand ideas about harm and conceptualizations of harassment.

While the authors are from different regions represented in the dataset, the project is centered at one institution in the United States. This means our perspectives are influenced by U.S.-centered approaches and values. Researchers based in other regions can contribute valuable perspectives by designing cross-cultural studies grounded in non-Western perspectives and justice frameworks \cite{pendse2021can}.

As mentioned in Section \ref{survey-design}, we selected a non-representative and non-comprehensive set of harassment scenarios and cannot fully explain variances in harm and response ratings for these scenarios. For example, while taking sexual photos without permission and sharing them on social media clearly elicited higher harm, we do not know how participants interpreted the prompt or what the specific kinds of harms they were imagining were.

Lastly, we acknowledge the use of frequentist statistics can lead to over- or mis-interpretation of results. We focused on prominent themes in our results which may be less prone to arise out of chance.

\section{Discussion}

\subsection{Women Perceive Greater Harm Than Men}
Our results show that women perceive greater harm associated with online harassment than men across many regions around the world. The scenario of taking sexual photos without permission and sharing them on social media was highest in harm for both women and men, and was higher for women than men. Advocates and activists globally have been fighting for greater protection of women's rights related to non-consensual sharing of sexual photos. In her book, \emph{Nobody's Victim}, United States-based lawyer Carrie Goldberg writes about non-consensual sharing, ``We are putting tech companies on notice. For too long, dating apps and other digital products have been enabling [people] to commit heinous crimes that put us all at risk. It's time these companies are held accountable. They need to think differently about the responsibility they owe us all.'' \cite{goldberg2019nobody}. 

These calls to action are needed across both corporations and policy-making. {S}cholar Nanjira Sambuli describes how an anti-pornography law in Uganda that could be intended to support victims may also be the law that can be used to punish women for images shared online without their consent \cite{challenge-africa-2015}. Mariana Valente, based in Brazil, has pointed out that misogynistic cultures in Brazil have led to criminal cases involving online defamation suits by men against women for calling them sexist, rather than cases that protect women \cite{internet-detox2019}. In South Korea, Team Flame, a team of women undergraduate students that publicized the Nth Room case---the cyber-trafficking case---led to greater sanctions for sharing non-consensual videos online {and revised laws for social platforms to curb online sexual violence} \cite{spark2020,sungwha2020}.
In the United States, Goldberg, along with legal scholars Danielle Citron and Mary Anne Franks, have been pivotal in pushing through greater regulatory protection to prevent non-consensual sexual image sharing \cite{goldberg2019nobody,keats2018sexual,citron2014criminalizing}. 

However, learning that men perceive lower harm than women in all four types of harassment, not just non-consensual sharing of sexual photos, across all regions studied, is concerning. {Scholars and technologists may perceive only some kinds of harassment as gendered (e.g. those that are explicitly gendered like sexist comments), but how people experience various interactions may vary based on their prior experiences or identities \cite{katz1996effects}. Importantly, interactions may not have to contain overt sexual language or expressions to cause harm.}  Many prominent technology companies are predominantly led by men and there may be a misalignment between how they believe women experience harm on their platforms versus how women see themselves experiencing harm. This misalignment may be especially difficult to detect when harms are difficult to measure; unlike some harms (e.g., theft of a bicycle, a broken arm), it remains difficult to quantify how online harassment harms its targets. 

A harm-based perspective may help social media companies to better address disparities on their platform. That is, platforms could adopt guiding frameworks, drawn from human rights and civil rights principles, that seek to recognize and repair harms. Such an approach could also help to address different experiences of harm in different regions. For example, in Colombia there have been widespread concerns and growing public awareness about violence against women. Hundreds of women are murdered per year on account of their gender in Colombia \cite{nowhere-colommbia} and as a result, Colombian women (and some men) have joined the online movement \#NiUnaMenos and \#NiUnaMás (\#NotOneLess and \#NotOneMore) to reflect on unequal power relationships that result in subordination and violence. 

These movements highlight how when we talk about online harassment and harm, we must consider the broader contexts in which those harms are taking place. Though harms can be difficult to quantify or measure, they also invite the perspectives of those who experience the harm and could enable us to imagine different kinds of responses to them.


\subsection{Exploring Justice Models {While Centering Those Most Affected by Harassment}}

Our results showed that participants are generally favorable towards most of the platform responses. That is, on average, they rate all of the responses as more desirable than undesirable. We intentionally chose a wide range of platform responses that reflect a range of justice frameworks. For example, apologies are aligned with restorative justice frameworks that seek to recognize and repair harm \cite{schoenebeck2020drawing}, whereas revealing identities publicly invokes a kind of vigilante justice that relies on public shaming \cite{nussbaum2009hiding}. Participants' general favorability towards all of these platform responses, despite the differences between them, suggests that---to some extent---they may be more eager to have \textit{any} kind of remedy rather than nothing. 

{W}omen prefer banning more than men in all four harassment scenarios. This may reflect differing frequency or severity of harassment that women experience online. Given that women experience various kinds of gender-based harm throughout their lives \cite{beres2007spontaneous}, banning users could be motivated by a sense of urgency to address and minimize any potential for future harm. Whereas an apology may be appropriate for less severe harassment, banning users is a more severe {sanction} and may be especially desirable for more severe kinds of online harassment.


Schoenebeck et al. \cite{schoenebeck2020drawing} argue that existing models of governance---namely, removing content and banning users---reflect criminal justice models that focus on punishing perpetrators of harassment. Building on arguments for criminal justice reform, they suggest that punitive models may fail to acknowledge the experience of targets or to hold perpetrators accountable for harms. Alternative justice theories, such as restorative justice, racial justice, or {even shame as a form of} justice may provide new avenues for how platforms respond to harassment \cite{schoenebeck2020drawing, hasinoff2020promise, sultana2021unmochon}. {While alternative justice theories could be compelling in some contexts, our findings show that they should be engaged with carefully and contextually when centering the voices of people who are harmed. In particular, our findings of women preferring banning---a punitive measure---more than men indicates women may prefer {more strict sanctions, especially in the absence of any other effective remedies}. Our results also indicate that though women were generally more favorable than men about most of the platform responses, they were less favorable towards payment as a response. 

Though future work should examine why, one possible interpretation is that punitive measures like banning meet victims' needs in ways which current restorative measures do not when their major concern is not reparation \cite{stubbs2007beyond}. 
Simultaneously, while both men and women preferred apologies less than banning, women preferred it more than men, indicating that nuanced responses may be needed depending on the nature of the harassment.
More research should examine what women's reasons are for these preferences and understanding the dynamics between punitive and restorative measures on social platforms.

} 

As many Internet scholars have argued, context matters. Differences in cultures and judicial systems across regions may impact women's preferences---for example, being a victim of gender-based online harassment might heighten the risk of reputational damage for the victim in regions with strong patriarchal or misogynistic values \cite{al2016exploratory, abokhodair2017photo, ahmad2009doesn, abu2016veiled, khan2008violence}.
We also speculate that {for some regions,} trust in judicial systems may shape people's preferences for responses. {In many countries, legal systems are criticized for taking sexual assault cases lightly, making it difficult for people who are harmed to find justice through them \cite{korea-change}.} This may cause people to prefer punitive and decisive responses existing on social platforms like removal and banning. 

Finally, a number of countries in our sample are post-colonial countries. While it is widely acknowledged that the colonial experience affected the political development of ex-colonial countries \cite{10.2307/20069916}, it may have also have helped shape punitive online governance preferences, as colonial justice was largely punitive in nature \cite{Kolsky2010ColonialJI, 10.2307/20069916}.

Ultimately, our results contribute to the conversations around the importance of centering those who are most affected by online harm in social media governance. Importantly, preferences may vary by cultures and individuals, suggesting that one-size-fits-all remedies may be ineffective or harmful in themselves  \cite{karusala2019privacy,sultana2018design,rabaan2021daughters,schoenebeck2020drawing,jiang2021understanding}. Though we focused on women's experiences, it is likely that these cumulative experiences of harm shape how targets in a variety of groups---non-binary people, transgender people, people of color in the United States, lower caste people in India, etc---may prefer to seek justice in their online experiences.

\subsection{Categories and Universality in Online Harassment}
Violence against women is a global cause embraced by the United Nations, World Health Organization, Amnesty International, and other organizations \cite{unesco-onlineviolence2021,humanrights-womenviolence-2019,aziz2015eliminating,un-takefive-2020}; however, it essentializes gender rather than recognizing the mutability and non-binary properties of gender \cite{scheuerman2019computers,bivens2016baking,keyes2018misgendering}. Gender binaries are social, economic, and political categories \cite{peterson1999political,hyde2019future}. In some of the countries we studied, it is illegal to be transgender or non-binary, and in many of them, it is not socially {or culturally} accepted (e.g., \cite{nova2021facebook}). We chose not to ask about non-binary identity or transgender identity because responding truthfully to those questions could put participants at risk in some regions; at the same time, overlooking them further marginalizes those {groups}. 

In our analysis, we also wrestled with the treatment of countries. We chose to analyze responses {in aggregate} to recognize universality of experiences and by country (plus the collection of countries in the Caribbean) in recognition of the cultural, social, economic, and political conditions of those regions. This was partly a conceptual and analytical choice---``global'' or ``non-Western'' studies often define populations within the boundaries of a country. It was also a methodological one---{country is a required selection criteria in panels}.

However, recognizing universality is a contested matter. In many ways universality may signify what Gramsci refers to as ``cultural hegemony,'' in which beliefs and value systems are mediated by those in power \cite{lears1985concept}. Anna Lauren Hoffmann, based in the United States, writes about the epistemologies and ideals associated with binary categories \cite{hoffmann2021even}. Hoffmann calls on readers to critically reflect on the idea of universalism which is often conflated with imperialism or the West \cite{hoffmann2021even}. Doing this does not mean giving up on universalism itself, but merely that what has been labeled universal is often far from it. In the context of women's experiences online, there may be some near-universal experiences of harm which reflect offline inequalities. However, there are also regional differences which reflect cultural values, policies, and laws that shape women's experiences of justice online. Addressing these requires that people and institutions in positions of power shift from aspirations of neutrality, towards more principled governance that centers the well-being of those who experience systematic harm.

\section{Conclusion}
This work reveals that on average, women perceive higher harm associated with online harassment than men do across 14 regions around the world. Perceived harm is highest for sharing of non-consensual sexual photos. It also reveals that on average, revealing and banning were the most preferred platform responses. When considering gender, women participants prefer banning users and apologies more than men, but they prefer payment less than men. Responses to harassment are generally desirable for all participants; however, women find most responses more desirable than men do. This work contributes to an expanding movement to decenter US perspectives on social media governance, while centering conversations about global and local values. It also draws attention to the limitations of principles like neutrality in content moderation if they are enacted in fundamentally inequitable social contexts.

\begin{acks}
We thank Anandita Aggarwal, Ting-Wei Chang, Chao-Yuan Cheng, Yoojin Choi, Banesa Hernandez, Kseniya Husak, Jessica Jamaica, Wafa Khan, and Nurfarihah Mirza Mustaheren for their contributions to this project. We thank Michaelanne Thomas, David Nemer, and Katy Pearce for early conversations about these ideas. We thank members of the Social Media Research Lab for their feedback at various stages of the project. We also thank the reviewers for their constructive comments. Im would especially like to thank Yoon Jeong Cha, Yoojin Choi, Song Mi Lee, and Wookjae Maeng for their advice and Shubham Atreja, Agrima Seth, and Anjali Singh for proofreading the paper. This material is based upon work supported by the National Science Foundation under Grants \#1763297 and \#1552503 and by a gift from Instagram. 

\end{acks}

\bibliographystyle{ACM-Reference-Format}
\bibliography{sample-base}

\received{April 2021}
\received[revised]{November 2021}
\received[accepted]{March 2022}

\end{document}